\documentclass{article}

\usepackage[final]{neurips_2023_ml4ps}

\usepackage[utf8]{inputenc} 
\usepackage[T1]{fontenc}    
\usepackage{hyperref}       
\usepackage{url}            
\usepackage{booktabs}       
\usepackage{amsfonts}       
\usepackage{nicefrac}       
\usepackage{microtype}      
\usepackage{xcolor}         
\usepackage{graphicx}

\title{JetLOV: Enhancing Jet Tree Tagging through Neural Network Learning of Optimal LundNet Variables}

\author{%
Mauricio A. Diaz$^1$ \quad Giorgio Cerro$^{1}\thanks{g.cerro@soton.ac.uk}$ \quad Jacan Chaplais$^2$ \\ \quad \textbf{Srinandan Dasmahapatra$^2$} \quad \textbf{Stefano Moretti$^{1,3}$}\\ \\
$^1$School of Physics and Astronomy, University of Southampton, Southampton, UK\\
$^2$School of Electronics and Computer Science, University of Southampton, Southampton, UK\\
$^3$Department of Physics and Astronomy, Uppsala University, Uppsala, Sweden\\
}

\begin{document}

\maketitle

\begin{abstract}
Machine learning has played a pivotal role in advancing physics, with deep learning notably contributing to solving complex classification problems such as jet tagging in the field of jet physics. In this experiment, we aim to harness the full potential of neural networks while acknowledging that, at times, we may lose sight of the underlying physics governing these models. Nevertheless, we demonstrate that we can achieve remarkable results obscuring physics knowledge and relying completely on the model's outcome. We introduce JetLOV, a composite comprising two models: a straightforward multilayer perceptron (MLP) and the well-established LundNet. Our study reveals that we can attain comparable jet tagging performance without relying on the pre-computed LundNet variables. Instead, we allow the network to autonomously learn an entirely new set of variables, devoid of a priori knowledge of the underlying physics. These findings hold promise, particularly in addressing the issue of model dependence, which can be mitigated through generalization and training on diverse data sets. The code and models are publicly available at \url{https://github.com/GiorgioCerro/jetlov}.

\end{abstract}

\section{Introduction}
Jet tagging constitutes a fundamental yet intricate aspect of jet physics. Its primary objective is the accurate identification of high-energy particles responsible for initiating cascades, ultimately leading to the formation of particle clusters known as jets. Within this context, it is imperative to distinguish between various types of particles that can trigger such events. However, the complexity is amplified when dealing with extreme energy regimes, making the task of correct identification exceedingly challenging.
Machine learning has made significant strides in aiding the physics community in improving models and achieving remarkable results in jet tagging. Some of the most recent state-of-the-art models, such as Particlenet \cite{Qu:2019gqs}, LundNet \cite{Dreyer:2020brq}, and ParticleTransformer (PartT) \cite{Qu:2022mxj}, rely on Graph Neural Networks. Our focus in this experiment centers on assessing the performance of LundNet, a model that has excelled not only due to its complexity and sophistication but also because of the pre processing of input data fed into the neural network.

LundNet utilises the Lund plane projection to transform the tree-like structure of jets into meaningful input features for each node in the tree \cite{Dreyer:2018nbf}. By considering the four-momentum of the two particle's descendants, LundNet computes a set of five variables known as LundNet variables. These variables encode essential information about the energy distribution and flow within the tree structure.
While it is often the case that more complex architectures yield superior classification performance, they often grapple with the issue of model independence. Recent findings, as presented at the BOOST 2023 workshop on behalf of the ATLAS collaboration \cite{MLeBlanc:2023}, have underscored the challenges of generalisation for intricate taggers. These models tend to be sensitive to the specifics of synthetic data generation, where variations in simulation software, such as PYTHIA \cite{Sjostrand:2014zea} and HERWING \cite{Bellm:2013hwb}, can significantly impact the outcomes; different softwares have their own algorithms for simulating physics processes, for example the parton shower and the hadronization steps.

Motivated by the quest for model independence in physics, we introduce JetLOV, a composite of two models which are first trained separately and then put together for further training on the jet tagging: RegNet (a Multilayer Perceptron network, MLP) is the one responsible for learning new set of variables to feed into the second part of the model, LundNet, which is responsible for the tagging. JetLOV aims to discover a new set of variables that can yield state-of-the-art performance. This approach paves the way for future endeavors where RegNet can be leveraged to learn variable sets that are independent of the data type provided. Our objective is to train the model to achieve model independence, offering a promising path forward in the realm of physics modeling. RegNet has a total of 55941
learnable parameters.

\section{Experiment}
The experiment is divided into two distinct steps. The initial step is dedicated to the training of the RegNet component (coded using Pytorch 1.7 \cite{paszke2019pytorch}) and involves a regression task aimed at learning the five LundNet variables for each particle denoted as "$k$" based on the four-momentum vectors of their two descendants, "$i$" and "$j$":

\begin{equation}
    (p_{\mu}^i, p_{\mu}^j) \rightarrow (\ln k_t, \ln \Delta, \ln z, \ln m, \psi) ^ k
\end{equation}

RegNet, which takes the form of a Multilayer Perceptron (MLP), is designed with five distinct branches, each responsible for independently learning one of the five variables. The branches handling "$\ln k_t$" and "$\ln m$" comprise a series of channels with dimensions (8, 128) and (128, 5), interconnected by Rectified Linear Unit (ReLU) activation functions. In contrast, the branches responsible for "$\ln \Delta$," "$\ln z$," and "$\psi$," which involve more intricate functions, employ a sequence of channels with dimensions (8, 128), (128, 128), and (128, 5), also incorporating ReLU activation functions for effective learning. 
This pre-training is performed in order to ensure that when we attach the second part of the architecture, which is already trained as well, we start not too far off the minimum.  Our investigation confirms the necessity of this pre-trained part for achieving high performance, as the non pre-trained RegNet fails to discover such optimal local minima.

In the second step we attach RegNet to the pre-trained LundNet model. In this way when we feed the data into the first model, it produce the input for the second, which then produce the final output, i.e. the probability for classifying the type jet. Once the prediction is compared with the target, we do the back propagation all the way back in order to update the weights of the full model. The goal is to minimise the Cross Entropy Loss and achieve a good performance. We look at five metrics: accuracy, area under the ROC curve (AUC), background rejection at three signal efficiencies (0.3, 0.5 and 0.7).  

\subsection{Data set}
The data set is taken from \cite{dataset}. In this project we worked only on the W-tagging, a standard binary classification problem, which consists of two classes: the signal (the W-jets, $pp \rightarrow WW$ process, and W required to decay hadronically) and the background (the QCD-jets, $pp \rightarrow jj$). Jets are clustered using the anti-kt algorithm \cite{Cacciari:2008gp}, \cite{Cacciari:2011ma} with a radius R = 1.0 using \textit{FastJet 3.3.2}, and are required to pass a selection cut, with transverse momentum $pt$ $>$ $500$ $GeV$ and rapidity $|y| < 2.5$. In each event, only the two jets with the highest transverse momentum are considered, and are saved as training data if they pass the selection cuts.

\section{Results}
\begin{figure}[h]
  \centering
    \includegraphics[width=14cm, height=9cm]{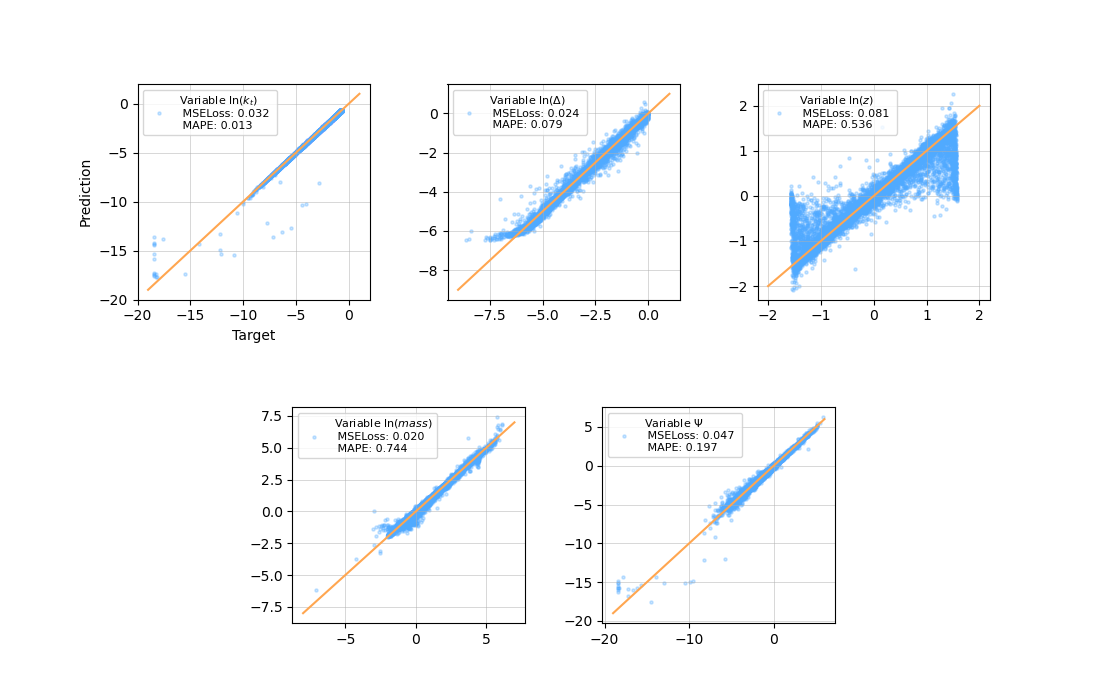}
  \caption{For each one of the LundNet variables we plot the prediction from the RegNet's outputs. We also report the Mean Squared Error (MSE) and the Mean Absolute Percentage Error (MAPE). The overall MSE for all the five variables is 0.037 on approximately 15000 points. Blu dots are the data points while the orange lines are the function $f(x) = x$.}
  \label{scatterplot}
\end{figure}

For the regression part, we trained the RegNet model with a data set of 100k events (50k signals and 50k backgrounds) and a smaller validation data set of 10k events. The training has been done over 100 epochs, until we reached a satisfying Mean Squared Error (MSE) Loss. In Fig. \ref{scatterplot} we can see the plots between target and prediction for each of the LundNet variables, the performance has been evaluated on a total number of particles of approximately 15000 and achieved a MSE of 0.037. All of them follow the expected linear trend, although it seems that the Neural Network struggles a bit on learning the third variable $\ln (z)$. 

For the classification, we have used an equally balanced data set of 1M events for the training, and 100k for the validation and 100k for the testing. The training, performed on a Nvidia GTX 1080 Ti graphics card with a minibatch size of 256, has been run for 30 epochs, with an initial learning rate of 0.001, and a scheduler that lowered the learning rate by a factor of 10 after the 20th and the 25th epochs. For the optimisation we have chosen to use Adam optimizer \cite{kingma2014adam}. The best model is chosen when the highest validation accuracy has been reached. In Tab. \ref{ROC}, we can see that the models achieve same performances, with slightly preference of LundNet on higher signal efficiency and a preference of JetLOV for lower signal efficiency.

\begin{table}[h]
  \caption{Results of several metrics for the two models (LundNet and JetLOV) for the W-tagging problem. The first column gives the area under the ROC curve, the second gives the accuracy, and the later three show the background rejection $(1/\epsilon_B)$ at three different signal efficiencies $(\epsilon_S)$, 30 $\%$, 50 $\%$ and 70 $\%$ respectively. For each metric, larger values indicate better performance.
}
\label{sample-table}
  \centering
  \begin{tabular}{lccccc}
    \toprule
    \multicolumn{2}{r}{AUC}  &  Acc. & 1/$\epsilon_B$ at $\epsilon_S$=0.3 &  1/$\epsilon_B$ at $\epsilon_S$=0.5 & 1/$\epsilon_B$ at $\epsilon_S$=0.7\\
    \midrule
    LundNet & 0.938 & 0.872 & 4545.4 & \textbf{602.4} & \textbf{70.2}\\
    JetLOV  & 0.938 & 0.872 & \textbf{6250.8} & 555.6 & 63.9\\
    \bottomrule
  \end{tabular}
  \label{ROC}
\end{table}

\begin{figure}[h]
  \centering
    \includegraphics[width=8cm, height=6cm]{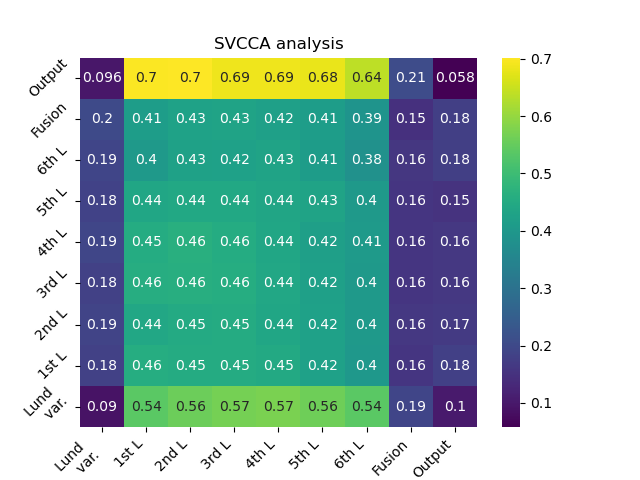}
  \caption{SVCCA analysis for each pair of the layers of the fully trained LundNet model before and after being combined with RegNet. Rows correspond to the model before being attached to RegNet, while cols correspond to the model after.}
  \label{svcca}
\end{figure}

After training, our primary focus shifted towards the learned variables from the initial part of the model that feed into the second part. We aimed to assess the extent to which these learned variables differed from the original LundNet variables. To conduct this analysis, we employed Canonical Correlation Analysis (CCA) to compare the original LundNet variables with the set learned by JetLOV.
The results of this analysis, conducted on a data set comprising 500 events, revealed a CCA value of 0.522, which indicates an absence of any significant correlation between the two sets of variables. Delving deeper, we further investigated the matter using Singular Vector Canonical Correlation Analysis (SVCCA) \cite{raghu2017svcca} between the output layers of the fully trained LundNet model before and after its attachment to RegNet. This comparative analysis is visualised in Fig. \ref{svcca}.
The SVCCA results provide compelling evidence that we have indeed identified a minimum point in our training process that satisfies the performance requirements for jet tagging. However, it is remarkable that the set of variables discovered at this minimum point deviates substantially from the original LundNet variables.
This intriguing outcome suggests that in certain cases, Machine Learning models can excel without relying on physics-derived features. While this may be advantageous as it hints at the potential for uncovering unconventional insights, there is a notable trade-off. The balance must be carefully struck between model interpretability and performance. Physics-based features are often interpretable, allowing researchers to comprehend the rationale behind a model's predictions. Conversely, when models learn novel features, they may become more opaque or 'black-box,' presenting challenges in interpreting their decision-making processes.

\section{Conclusion}

Jet tagging represents a critical and intricate task within the domain of jet physics. Its successful execution relies on accurately identifying high-energy particles that initiate cascades, ultimately forming clusters of final-state particles known as jets. While recent state-of-the-art models exhibit exceptional performance on specific tasks, they grapple with a challenge: model independence. These models' performance varies significantly based on the algorithms used for specific physics processes (such as parton shower and hadronization) and the choice of software for data generation.

Motivated by the imperative for model generality, we introduce JetLOV—an ensemble comprising two neural networks. JetLOV demonstrates promising performance in W-tagging, a task involving the classification of W jets versus QCD jets, when compared to the established LundNet model. The unique design of JetLOV entails two stages: first, RegNet, an MLP, learns to reproduce the LundNet variables. Subsequently, when integrated with the pre-trained LundNet, JetLOV is further trained on a full data set of 1 million events, aiming to discover an alternative minimum. The results are striking: JetLOV successfully identifies another minimum, effectively generating a distinct set of "LundNet variables" that match the performance of the original LundNet, albeit with a slight preference for higher background rejection at the cost of slightly reduced signal efficiency. This result offers a compelling example of how machine learning can diverge from traditional physics yet yield exceptional outcomes. It underscores the delicate balance between model interpretability and performance, opening new horizons for advancing neural network taggers.

While this preliminary experiment showcases the potential of our approach, our primary focus for future research lies in augmenting machine learning models with universality. To this end, we plan to replicate our experiment using data sets from diverse sources, including those generated with different setups. This endeavour aims to address the critical challenge of model independence more comprehensively, enabling ML models to transcend the constraints of specific data generation methods and achieve broader applicability in the field of physics.

\section*{Acknowledgements}
We acknowledge the use of the IRIDIS High Performance Computing Facility, and associated support services, at the University of Southampton, in the completion of this
work.
\bibliographystyle{plain} 
\bibliography{biblio.bib}

\newpage

\section*{Checklist}

\begin{enumerate}

\item For all authors...
\begin{enumerate}
  \item Do the main claims made in the abstract and introduction accurately reflect the paper's contributions and scope? YES
  \item Did you describe the limitations of your work? YES
  \item Did you discuss any potential negative societal impacts of your work? YES
  \item Have you read the ethics review guidelines and ensured that your paper conforms to them? YES
\end{enumerate}

\item If you are including theoretical results...
\begin{enumerate}
  \item Did you state the full set of assumptions of all theoretical results? N/A
        \item Did you include complete proofs of all theoretical results? N/A
\end{enumerate}

\item If you ran experiments...
\begin{enumerate}
  \item Did you include the code, data, and instructions needed to reproduce the main experimental results (either in the supplemental material or as a URL)? YES
  \item Did you specify all the training details (e.g., data splits, hyperparameters, how they were chosen)? YES
        \item Did you report error bars (e.g., with respect to the random seed after running experiments multiple times)? N/A
        \item Did you include the total amount of compute and the type of resources used (e.g., type of GPUs, internal cluster, or cloud provider)? YES 
\end{enumerate}

\item If you are using existing assets (e.g., code, data, models) or curating/releasing new assets...
\begin{enumerate}
  \item If your work uses existing assets, did you cite the creators? N/A
  \item Did you mention the license of the assets? N/A
  \item Did you include any new assets either in the supplemental material or as a URL? N/A
  \item Did you discuss whether and how consent was obtained from people whose data you're using/curating? N/A
  \item Did you discuss whether the data you are using/curating contains personally identifiable information or offensive content? N/A
\end{enumerate}

\item If you used crowdsourcing or conducted research with human subjects...
\begin{enumerate}
  \item Did you include the full text of instructions given to participants and screenshots, if applicable? N/A
  \item Did you describe any potential participant risks, with links to Institutional Review Board (IRB) approvals, if applicable? N/A
  \item Did you include the estimated hourly wage paid to participants and the total amount spent on participant compensation? N/A
\end{enumerate}

\end{enumerate}

\end{document}